\documentclass [11pt,a4paper]{article}

\usepackage{amsmath}
\usepackage{amsthm}
\usepackage{color,graphicx}
\usepackage{latexsym}
\usepackage{amssymb}
\usepackage{enumerate}
\usepackage[latin2]{inputenc}
\usepackage{color}
\usepackage{pslatex}

\newtheorem* {theorem*}      {Theorem}
\newtheorem* {lemma*}        {Lemma}
\newtheorem* {corollary*}    {Corollary}
\newtheorem* {proposition*}  {Proposition}
\newtheorem* {definition*}   {Definition}
\newtheorem* {remark*}       {Remark}

\newtheorem* {remarks*}      {Remarks}

\newtheorem* {claim*}        {Claim}

\numberwithin{equation}{section}

\numberwithin{figure}{section}

\def \N {\mathbb N}

\def\beqs{\begin{eqnarray*}}
\def\eeqs{\end{eqnarray*}}
\def\beq{\begin{eqnarray}}
\def\eeq{\end{eqnarray}}
\def\beas{\begin{eqnarray*}}
\def\eeas{\end{eqnarray*}}
\def\bea{\begin{eqnarray}}
\def\eea{\end{eqnarray}}

\def \prob        {\ensuremath{\mathbb{P}}}
\def \expect      {\ensuremath{\mathbb{E}}}
\def \indicator   {\ensuremath{\mathbb{I}}}

\def \cikk        {Note}
\def \AMFP        {A.M.F.I.P.}
\def \ATSP        {A.T.S.P.}
\def \ER          {Erd\H{o}s-R\'enyi }

\def \rr          {\mathbf{r}}
\def \xx          {\mathbf{x}}
\def \vv          {\mathbf{v}}
\def \XX          {\hat{X}}
\def \VV          {\hat{V}}
\def \ww          {\mathbf{w}}

\def \mm          {m_1^*(t)}
\def \mmn         {m_1^*(0)}

\newcommand{\abs}[1]{\left|{#1}\right|}

\newcommand{\CC}       {\ensuremath{\mathcal C}}
\def \Odd              {\mathbb O}
\def \Oddinf           {{\mathbb O}^\infty}
\def \sizeofgiant      {v_{\infty}(t)}

\title{Triangle percolation \\ in mean field random graphs --- with PDE}

\author{
{\sc Bal\'azs R\'ath} \qquad and  \qquad {\sc  B\'alint T\'oth}
\\[5pt]
Institute of Mathematics
\\
Budapest University of Technology (BME)
}

\begin{document}

\maketitle

\begin{abstract}
We apply a  PDE-based method to deduce the critical time
and the size of the giant component of the ``triangle percolation'' on
the \ER random graph process investigated by Palla, Der\'enyi and
Vicsek in  \cite{VicsekPhysRevLetters}, \cite{VicsekMasik}.
\end{abstract}

\section{Introduction}
In this \cikk  \text{ }we investigate triangle percolation in the
\ER random graph and other related graphs. The model  was defined
in \cite{VicsekPhysRevLetters} and \cite{VicsekMasik} by Derényi,
Palla, and Vicsek  in order to simulate phase transition
 of overlapping communities in real networks. They also considered
the more general case of $k$-clique percolation, but we restrict
our attention to the $k=3$ case.

The \ER random graph $G(N,p)$, defined in \cite{ErdosRenyi},  is a
random subgraph of $K_N$, the complete graph on $N$ vertices: The
edge set of  $G(N,p)$  is chosen at random in the following
manner: we declare every edge occupied with probability $p$,
otherwise we call the edge vacant. If $N$ is very large and
$p=\frac{t}{N}$, it is well-known that the random graph undergoes
phase transition: if we consider the size distribution of the
connected components, then a giant component will emerge at $t_c=1$
(this is called the critical time in the $k=2$ case). Because of
the mean field property of the graph, it is possible to relate the
distribution of the size of the connected component of an
arbitrary vertex to the total population of a Galton-Watson
branching process with a $\lambda =t$ parameter Poisson offspring
distribution (see e.g. \cite{JansonLuczakRucinski}).

 The branching process (and in the $N \to \infty$ limit, the
random graph) becomes supercritical when $t$, the expected number
of first generation offspring (i.e. the number of neighbors of the
root) exceeds $1$. The
limiting component-size distribution of the random graph and the
density of the giant component can be determined explicitly with
the generating function method.

Vicsek et al. generalize this idea to determine the critical time
of a different type of phase transition: if we consider the
triangle subgraphs of the \ER random graph and declare two
triangles connected if they have a common edge, then a different
time-scale is needed to see the emergence of a connected triangulated
giant
component: the branching process method makes sense in the $N \to
\infty$ limit if $p= \frac{t}{\sqrt{N}}$,  and
$t_c=\frac{1}{\sqrt{2}}$ is the critical time of the $k=3$ case, see
\cite{VicsekMasik}.

A different approach to arrive at the same results (in the $k=2$
case) is to view the evolution of the random graph as a stochastic
process: if every edge of the graph turns from empty to occupied state
with rate
$\frac{1}{N}$, independently from one-another then at time $t$ we will
see
$G(N,p=1-e^{-\frac{t}{N}})$ which is asymptotically the same as
$G(N,p=\frac{t}{N})$ as $N \to \infty$. It is easy to relate the
evolution of the random graph to the mean field stochastic model
of coagulation, the Marcus-Lushnikov process, which converges to
the solution of the Smoluchowski coagulation equation (with
multiplicative kernel), see (in historical order)
\cite{Buffet_Pule_1}, \cite{Buffet_Pule_2}, \cite{Aldous},
\cite{Fournier}.  A
useful way to handle the Smoluchowski equation is by taking its
Laplace-transform (which is essentially the same as using
generating functions): the transformed differential equation
becomes a well-known PDE, the Burgers equation (see
\cite{Buffet_Pule_1}, \cite{Buffet_Pule_2},
\cite{MenonPego}), which can be solved explicitly by using the
method of characteristics.

The method of branching processes cannot be applied to a slightly
modified random graph process, where the dynamics are the same but
the initial graph is an arbitrary graph (and not the empty graph
as in the \ER case), but the PDE method generalizes to these
models both in the $k=2$ and the $k=3$ case. In this \cikk{} we
derive a (non-linear, non time-homogeneous, first order) PDE which
describes the evolution of the $N \to \infty$ limit of the
triangulated component-size densities of a general mean-field
random graph, and give an explicit solution to that PDE, which
enables us to calculate the critical time and the size of the
giant triangulated component for an arbitrary initial component
size-distribution.

\section{Definitions}

In the rest of this \cikk, we will look at the random graph
processes on the following time-scale: $G(N,p \approx
\frac{t}{\sqrt{N}})$, or more briefly $G(N,t)$: vacant edges
turn occupied independently with rate $\frac{1}{\sqrt{N}}$. The
edge set of $G(N,t)$ is denoted by $E(N,t)$. In the \ER case, the
initial graph has no edges, but it can be itself a random graph in the
general case.

In order to describe the triangle-structure of the graph, let us
define an auxiliary graph, $\hat{G}(N,t)$ with vertex set
$E(N,t)$, and two vertices of $\hat{G}(N,t)$ be connected by an
edge if the corresponding $e, f \in E(N,t)$ are edges of the same
triangle whose third edge is also in $E(N,t)$. We call the
connected components of $\hat{G}$ the triangulated components of
$G$. The ``size'' or ``weight'' of a triangulated
 component is its size in $\hat{G}$.

If $e \in E(N,t)$, denote by $S(e,t)$ the size of the triangulated
component of $e$.
 Let us denote by ${\CC}_n(N,t)$ the number
 of triangulated components of weight $n$. Summing the total weight of
 components we get the total number of
 edges: $\sum_{n\in \N} n \cdot {\CC}_n(N,t) = \abs{E(N,t)}$,
 and we can define
\begin{equation}\label{elekszama}
\mm:=\lim_{N \to \infty} \frac {\abs{E(N,t)}} {\frac{1}{2}
N^{\frac{3}{2}}} =\mmn+t
\end{equation}
by the law of large numbers.

 If we define
  $c_n(N,t)= \frac{{\CC}_n(N,t)}{\frac{1}{2} N^{\frac{3}{2}}}$, then
  it is natural to expect that  $\lim_{N \to \infty} c_n(N,t)=c_n(t)$
  exists and is a deterministic nonnegative real number for each $n$ and
  $t$. Of course, there is a minimal criterion for this to hold:
   the sequence of initial
   random graphs must have the property that the limits
$\lim_{N \to \infty} c_n(N,0)=c_n(0)$
  exist. We need additional assumptions:

The \emph{asymptotic mean field independence property}, or briefly
\AMFP: Let us choose a  subset $E' \subseteq E(N,t)$ with
$\abs{E'} \ll N^{\frac{3}{2}}$ (for example $E'$ can be the set of
edges connected to $v_0 \in V(G)$), explore the connected
components of the edges of $E'$ in $\hat{G}(N,t)$ and denote the
edges contained in the explored triangles by $\overline{E'}$. The
\AMFP{ } is satisfied if the probability distribution of $S(f,t)$,
where $f \notin \overline{E'}$
 is asymptotically independent from the distribution of the explored
components as $N \to \infty$. Note that this property is weaker
then the "branching process" property:
 there asymptotic
independence holds even after a smaller exploration step: if $e
\in E'$, then the number of triangles that contain $e$, but are
not contained in $E'$ is asymptotically independent of $E'$.

We also assume that our sequence of mean field initial graphs has
the \emph{asymptotic trivial structure property}, or briefly
\ATSP: let us define another auxiliary graph, $\tilde{G}(N,t)$.
$V(\tilde{G}(N,t))$ consists of the triangles of $G(N,t)$ and two
vertices are connected if the corresponding triangles share an
edge. The \ATSP $ $ means that asymptotically almost surely the
connected component of an arbitrarily chosen vertex of
$\tilde{G}(N,t)$ is either the (unique) giant component of
$\tilde{G}(N,t)$ or a tree.

%The dynamics of our random graph process does not change the mean
%field property of the initial graph, so we assume that the \AMFP $
%$ and the \ATSP $ $ holds for all $t$.

An immediate consequence of the \ATSP $ $ assumption is that
$c_n(t)>0$ only if $n$ is in $\Odd$, the
 set of positive odd numbers: if we remove a vertex from a tree in
 $\tilde{G}(N,t)$, then we remove exactly two edges of $E(N,t)$.
  Denote by $\Oddinf=\Odd \cup \{ \infty
 \}$: in the mean field limit, the size of the triangulated
 component of an edge $e$ is in $\Oddinf$, $S(e,t)= \infty$ means
 that $e$ is an edge of the triangulated giant component.

Let us define  the Laplace transform (generating function)
 $C(t,x)=\sum_{n \in \Odd} c_n(t) \cdot e^{-n\cdot x}$  for $t \geq
 0$ and $x > 0$. $C(t,x)=\sum_{n \in \Oddinf} c_n(t) \cdot e^{-n\cdot x}$,
 since $e^{-\infty \cdot x}=0$.
  Denote the partial derivatives of $C$ with
 respect to $t$ and $x$ by $\dot{C}$ and $C'$, respectively.
 It is convenient to define $v_n(t)=n \cdot c_n(t)$ and
 $V(t,x)=\sum_{n \in \Odd} v_n(t) \cdot e^{-n\cdot x}$, so that
 $C'(t,x)=-V(t,x)$ holds. Under the assumption of the mean field properties,
  the function $C$ will satisfy the following PDE:
\begin{equation}\label{vicsekmodellPDEC}
 \dot{C}(t,x)=e^{ V(t,x)^2- \mm^2 -x} -2 V(t,x) \cdot \mm^2
\end{equation}

After solving the PDE, we will be able to express $c_n(t)$ as a
function of the initial data.
 Let us emphasize that $m_1(t):=\sum_{n \in \Odd}
v_n(t)=V(t,0_+) \neq \mm$ for all $t$ (although we do assume that
$m_1(0)=\mmn$), because the $N \to \infty$ limit and the $\sum_{n
\in \Odd}$ summation are not interchangeable in the supercritical
phase: the breakdown of equality indicates the presence of a giant
component, because a positive portion of edges is missing if we
sum the weight of small components:
\begin{equation}\label{elek_a_veges_es_orias_komponensben}
  m_1(t)+\sizeofgiant=\mm=m_1(0)+t
\end{equation} where $\sizeofgiant \cdot \frac{1}{2} N^{\frac{3}{2}}$
is, up to leading order,  the
weight of the giant component of $\hat{G}(N,t)$.

\section{Derivation of the PDE}

In order to derive (\ref{vicsekmodellPDEC}), we need some more
definitions.

 Let us orient the edges of $K_N$ in an arbitrary way, so that we
can talk about the ``initial'' and ``final'' endpoints of each
edge. If $e$ is the new edge that we are about to occupy at time
$t$, with initial and final endpoints $u$ and $v$, then the
``vicinity'' of $e$  can be described by a two-variable function
$\rho_e^t: \Oddinf \times \Oddinf \to \N $, where $\rho_e^t(i,j)$
is the number of vertices $w$ such that $w$, $u$ and $v$ form an
``$(i,j)$-type cherry'': both $\{u,w\}$ and $\{v,w\}$ are  in
$E(N,t)$, moreover $S(\{u,w\},t)=i$ and $S(\{v,w\},t)=j$.
$i=\infty$ or $j=\infty$ is an admissible choice, because we want
to take into account those edges in the vicinity of $e$ that
belong to the giant triangulated component.
 When $e$ becomes occupied, all
the components in the vicinity of $e$ merge into one component of
size
 $1+\sum_{i,j} \rho_e^t(i,j) \cdot (i+j)$,  since the
 merged non-giant components are distinct by the \ATSP{} The value of
 $\CC_n(t)$
 changes by

\[\indicator \lbrack \text{ } n= 1+ \sum_{i,j} \rho_e^t(i,j) \cdot
(i+j)\text{ } \rbrack
 -\sum_i \rho_e^t(i,n) - \sum_j \rho_e^t(n,j) \]

 We can give the probability
distribution of $\rho_e^t$ for an arbitrary $e$ when $N \to
\infty$ using the \AMFP{} For each $w$,
\[ \prob(S(\{u,w\},t)=i) \approx \frac{i \cdot
  \CC_i(N,t)}{\binom{N}{2}} = \frac{v_i(t)}{\sqrt{N}}, \]
up to leading order. Also $\prob(S(\{u,w\},t)=\infty)\approx
\frac{v_{\infty}(t)}{\sqrt{N}}$. Using similar estimates and the
\AMFP, the probability that $w$, $u$ and $v$ form an $(i,j)$-type
cherry is $\frac{v_i(t) \cdot v_j(t)}{N}$, up to leading order.
When $N \to \infty$,  the number of $(i,j)$-type cherries in the
vicinity of $e$ has Poisson distribution and their joint
distribution is the product measure: for any fixed $\rr: \Oddinf
\times \Oddinf \to \N $,
 the probability of the event $\{ \forall i \forall j $ $
 \rho_e^t(i,j) = \rr(i,j) \}$ (or briefly $\{ \rho_e^t \equiv
 \rr  \}$) is

\[
  \prod_{i,j \in \Oddinf} e^{-v_i(t) v_j(t)}\frac{(v_i(t)
  v_j(t))^{\rr(i,j)}}{\rr(i,j)!}=
e^{-\mm^2} \prod_{i,j \in \Oddinf} \frac{(v_i(t)
v_j(t))^{\rr(i,j)}}{\rr(i,j)!}
\]

We can now start to derive the differential equation
(\ref{vicsekmodellPDEC}).

Between $t$ and $t+ \mathrm{d}t$, approximately $\frac{
  N^{\frac{3}{2}} }{2} \mathrm{d}t$ edges become occupied, and their
contributions to the change of $\CC_n(N,t)$ are independent again
by the \AMFP, so we may use the  law of large
numbers to describe the evolution of the component-size vector:
\begin{multline*}
 \CC_n(N,t+\mathrm{d}t) - \CC_n(N,t) \approx
 \expect \left( \CC_n(N,t+\mathrm{d}t) - \CC_n(N,t) \right) \approx\\
\sum_\rr \left( \indicator \lbrack 1+ \sum_{i,j} \rr(i,j) \cdot
 (i+j)=n \rbrack
 -\sum_i \rr(i,n) - \sum_j \rr(n,j) \right) \cdot \prob(\rho_e^t
 \equiv \rr)
\frac{ N^{\frac{3}{2}} }{2}\mathrm{d}t
\end{multline*}

If we divide both sides by $ \frac{ N^{\frac{3}{2}} }{2}
\mathrm{d}t$, let $N \to \infty$ and $\mathrm{d}t \to 0$  and take
the Laplace-transform of both sides, then the left-hand side
becomes $\dot{C}(t,x)$. Let us calculate the Laplace-transform of
the right-hand-side. The first term is
\begin{multline*}
\sum_{n \in \Oddinf} e^{-n \cdot x}
 \sum_\rr  \indicator \lbrack 1+ \sum_{i,j} \rr(i,j) \cdot (i+j)=n
 \rbrack
 \cdot \prob(\rho_e^t \equiv \rr) =\\
\sum_\rr \prob(\rho_e^t \equiv \rr) e^{ -\left(1 +\sum_{i,j}
 \rr(i,j)\cdot (i+j) \right) \cdot x} =
e^{-\mm^2 -x} \sum_\rr  \prod_{i,j} \frac{  \left(  (v_i(t)
    e^{-ix})(v_j(t) e^{-jx})  \right)^{\rr(i,j)}  }{\rr(i,j)!}=\\
e^{-\mm^2 -x} \prod_{i,j} \sum_{r=0}^{\infty}  \frac{  \left(
(v_i(t)
    e^{-ix})(v_j(t) e^{-jx})  \right)^r  }{r!}=
e^{-\mm^2 -x} \prod_{i,j}e^{(v_i(t) e^{-ix})(v_j(t) e^{-jx})}=\\
e^{V(t,x)^2-\mm^2-x}
\end{multline*}
The second term:
\begin{multline*}
\sum_{n\in \Oddinf} e^{-n \cdot x} \sum_\rr \sum_{i \in \Oddinf}
\rr(i,n) \prob(\rho_e^t \equiv \rr)= \sum_{n \in \Odd} e^{-n \cdot
x} \sum_{i \in \Oddinf} \expect
(\rho_e^t(i,n))=\\
\sum_{n \in \Odd} e^{-n \cdot x}  \sum_{i \in \Oddinf} v_i(t)
v_n(t)= \mm V(t,x)
\end{multline*}
The third term is handled in the same way.

Putting these equations together we arrive at
(\ref{vicsekmodellPDEC}). It is convenient to use the shorthand
notation $W(t,x):=e^{V(t,x)^2-\mm^2-x}$ for  the Laplace transform
(generating function) of the size of the component we get by
occupying a  vacant edge at time $t$. Note that $W(t,0_+)<1$
indicates that this probability distribution is defective in the
supercritical case, since $\mm- V(t,0_+)=\sizeofgiant>0$.

\section{Solution of the PDE}

It is possible to give an explicit solution to
(\ref{vicsekmodellPDEC}) with the
 method of characteristics. Differentiating the PDE with respect to
 $x$ and
 rearranging the equation we get a first order PDE for $V$:
% Rearranging the equation and substituting
%$V(t,0)=V(0,0)+t$ we get
\begin{equation}\label{vicsekPDErearranged}
 \dot{V}+V' \cdot \left( e^{V^2-\mm^2-x} \cdot 2V-2\mm
\right) =e^{V^2-\mm^2 -x}
\end{equation}
%The method of characteristics is based on the following idea: imagine
%that we already have $V(t,x)$ (the solution of the PDE) at hand,
%define a system
%of ordinary differential equations using $V(t,x)$, show that in fact
%the ODE
%can be written in a form that doesn't contain $V(t,x)$, solve the
%ODE,
%construct the solution of the PDE using the solution of the ODE,
%verify the solution.

Let us consider the following ODE with initial condition $\xx(0)=x$:

\begin{equation}\label{ODEx}
 \dot{\xx}(t)=e^{V(t,\xx(t))^2-\mm^2- \xx(t)
  }
\cdot 2 V(t, \xx(t)) -2 \mm
\end{equation}

If we define $\vv(t)=V(t,\xx(t))$, then
$\vv(0)=V(0,x)$. Putting (\ref{vicsekPDErearranged}) and
(\ref{ODEx}) together we get a system of differential
equations that can be solved without knowing $V(t,x)$ in advance:

\[
\begin{cases}
 \dot{\xx}(t)=e^{\vv(t)^2-\mm^2-\xx(t)}\cdot
2 \vv(t) -2\mm \\
 \dot{\vv}(t)=
e^{\vv(t)^2-\mm^2 - \xx(t)}
\end{cases} \]

In order to solve these equations explicitly,
 define $\ww(t)= e^{\vv(t)^2-\mm^2 -
   \xx(t)}=W(t, \xx(t))$.
\[
\dot{\ww}(t)=\ww(t) \cdot
 \left(
   2\vv(t)\dot{\vv}(t)-2\mm-\dot{\xx}(t)
 \right)%=\\
%\ww(t) \cdot \left( 2\vv(t)\ww(t)-2\mm -
%  (\ww(t) 2 \vv(t) -2\mm \right)
=0
\]
Thus $\ww(t)$ is constant:
$W(t,\xx(t))=W(0,x)$, $\vv(t)$ is linear:
$\vv(t)=V(0,x)+t \cdot W(0,x)$, and $\xx(t)$ is quadratic:
\begin{equation}\label{quadratic}
\xx(t)= x+ (V(0,x)+t \cdot W(0,x))^2 -V(0,x)^2-(m_1(0)+t)^2 +
m_1(0)^2 \end{equation}

If we start with $\xx(0)=0$, then $\xx(t) \equiv 0$ and $\vv(t)=
m_1(0)+t=\mm$, but we know that $V(t,0_+) \neq \mm$ if $t>T_g$.
This breakdown of analiticity is due to the intersection of
characteristics:
 another characteristic curve $\xx(t)$ starting at $x$
intersects the  $\xx(t) \equiv 0$ curve at the  time when
(\ref{quadratic}) becomes zero: the intersection time $t$ solves
the following equation:
\[
t^2 \cdot \left( W(0,x)^2 -1\right)+t \cdot \left( 2V(0,x)
W(0,x)-2m_1(0) \right) +x=0
\]

If we let $x \to 0$ in this equation, the solution will converge
to $t=T_g$, the first time when another characteristic curve hits
$0$. We have to divide all the coefficients by $x$, use
$m_1(0)=\mmn=V(0,0)= \sum_{n \in \Odd} v_n(0)$, $m_2(0)=-V'(0,0)=
\sum_{n \in \Odd} n\cdot v_n(0)$ and $W(0,0)=1$ to get $-2$ times
the following equation as $x \to 0$:
\begin{equation}\label{masodfokuegyenlet} T_g^2 \cdot (2m_2(0)m_1(0)+1) + T_g
\cdot \left( m_2(0) + m_1(0)\cdot (2m_2(0)m_1(0)+1) \right)
-\frac{1}{2} =0
\end{equation}

As a special case, if we start from the empty graph, then $m_1(0)
= 0$ and $m_2(0)=0$, thus we get $T_g=\frac{1}{\sqrt{2}}$, which
agrees with the critical time obtained in
\cite{VicsekPhysRevLetters}  and  \cite{VicsekMasik}, by using the
branching process method. If $G(N,0)$ is uniformly chosen from all
graphs with $m_1(0)\frac12 N^{\frac32}$ edges (where
$m_1(0)<\frac{1}{\sqrt{2}}$), then this graph is asymptotically
the same as the \ER graph at time $t=m_1(0)$,
 thus $m_2(0)=\frac{m_1(0)}{1-2
m_1(0)^2}$ and $T_g=\frac{1}{\sqrt{2}}-m_1(0)$. This result can
also be obtained by the branching process method.

 An example of a sequence of initial
random graphs that have the \AMFP,  but do not have the branching
process property: let $G(N,0)$ be chosen uniformly from all
triangle-free graphs that have $m_1(0)\frac12 N^{\frac32}$ edges.
In this case $m_2(0)=m_1(0)$, and the $T_g$ of this graph is
greater than the $T_g$ of the \ER graph with the same $m_1(0)$,
but smaller than that of the \ER graph with the same $m_2(0)$.
This follows from the fact that the solution of the equation
(\ref{masodfokuegyenlet}) decreases if we increase $m_1(0)$ or
$m_2(0)$.

In order to express the value of $\sizeofgiant$, let us define
$\XX(t,w)$, the inverse function of $W(t,x)$ in the $x$ variable
and $\VV(t,w)=V(t,\XX(t,w))$. $\XX(t,w)$  is well-defined and is a
decreasing function of $w$ on the interval $( 0, W(t,0) \rbrack $.
Since $w(t)$ remains constant along the characteristics,
$\VV(t,w)=\VV(0,w)+tw$ and
\[ \XX(t,w)=\XX(0,w)+(\VV(0,w)+tw)^2-\VV(0,w)^2-(m_1(0)+t)^2 + m_1(0)^2 \]
is expressed explicitly given the initial data.
 $W(t,0)$ is the
smallest $w$ such that $\XX(t,w)=0$, and $\sizeofgiant= \mm -
m_1(t)=\VV(0,1)+t-\VV(0,W(t,0))-tW(t,0)$.

\vskip1cm
\noindent
{\bf Acknowledgement:}
The research work of
the authors is partially supported by the following OTKA (Hungarian
National Research Fund) grants:
K 60708 (for B.R. and B.T.),
TS 49835 (for B.R.).

\bibliography{bib_rb}

\begin{thebibliography}{1}

\bibitem{Aldous}
D.~J. Aldous.
\newblock Deterministic and stochastic models for coalescence (aggregation and
  coagulation): a review of the mean-field theory for probabilists.
\newblock {\em Bernoulli}, 5(1):3--48, 1999.

\bibitem{Buffet_Pule_1}
E.~Buffet and J.V. Pul\`e.
\newblock On {L}ushnikov's model of gelation.
\newblock {\em Journ. Stat. Phys.}, 58:1041--1058, 1990.

\bibitem{Buffet_Pule_2}
E.~Buffet and J.V. Pul\`e.
\newblock Polymers and random graphs.
\newblock {\em Journ. Stat. Phys.}, 64:87--110, 1991.

\bibitem{VicsekPhysRevLetters}
I.~Der\'{e}nyi, G.~Palla, and T.~Vicsek.
\newblock Clique percolation in random networks.
\newblock {\em Phys Rev. Lett}, 94:160202, 2005.

\bibitem{ErdosRenyi}
P.~Erd{\H{o}}s and A.~R{\'e}nyi.
\newblock On the evolution of random graphs.
\newblock {\em Bull. Inst. Internat. Statist.}, 38:343--347, 1961.

\bibitem{Fournier}
N.~Fournier and J-S. Giet.
\newblock Convergence of the {M}arcus-{L}ushnikov process.
\newblock {\em Methodol. Comput. Appl. Probab.}, 6(2):219--231, 2004.

\bibitem{JansonLuczakRucinski}
S.~Janson, T.~{\L}uczak, and A.~Ru\'cinski.
\newblock {\em Random {G}raphs}.
\newblock Wiley-Interscience Series in Discrete Mathematics and Optimization.
  Wiley-Interscience, New York, 2000.

\bibitem{MenonPego}
G.~Menon and R.~L. Pego.
\newblock Approach to self-similarity in {S}moluchowski's coagulation
  equations.
\newblock {\em Comm. Pure Appl. Math.}, 57(9):1197--1232, 2004.

\bibitem{VicsekMasik}
G.~Palla, I.~Der\'{e}nyi, and T.~Vicsek.
\newblock The critical point of k-clique percolation in the
  {E}rd\"{o}s-{R}\'{e}nyi graph.
\newblock {\em J. Stat. Phys.}, 128:219--227, 2007.

\end{thebibliography}
\bibliographystyle{plain}

\vfill
\hbox{
\phantom{.}\hskip6cm
\vbox{\hsize=6cm
\noindent
Address of authors:
\\[5pt]
Institute of Mathematics
\\
Budapest University of Technology
\\
Egry J\'ozsef u. 1
\\
H-1111 Budapest, Hungary
\\[5pt]
e-mail:
\\
{\tt rathb{@}math.bme.hu\\balint{@}math.bme.hu}
}}

\end{document}